\documentclass[twocolumn,times]{aastex63}

\usepackage{newtxtext,newtxmath}
\usepackage{graphicx}	% Including figure files
\usepackage{amsmath}	% Advanced maths commands
\usepackage{amssymb}	% Extra maths symbols
\usepackage{mathrsfs}

\received{}
\revised{}
\accepted{}
\submitjournal{ApJL}

\shorttitle{Binary Comb Model for Periodic FRBs}
\shortauthors{Ioka \& Zhang}
\begin{document}

\title{A Binary Comb Model for Periodic Fast Radio Bursts}

\correspondingauthor{Kunihito Ioka}
\email{kunihito.ioka@yukawa.kyoto-u.ac.jp}

\author[0000-0002-3517-1956]{Kunihito Ioka}
\affiliation{Center for Gravitational Physics, Yukawa Institute for Theoretical Physics, Kyoto University, Kyoto 606-8502, Japan}

%\collaboration{1}{(AAS Journals Data Scientists collaboration)}

\author{Bing Zhang}
\affiliation{Center for Gravitational Physics, Yukawa Institute for Theoretical Physics, Kyoto University, Kyoto 606-8502, Japan}
\affiliation{Department of Physics and Astronomy, University of Nevada Las Vegas, Las Vegas, NV 89154, USA}



\begin{abstract}
We show that the periodic FRB 180916.J0158+65 can be interpreted by invoking an interacting neutron star binary system with an orbital period of $\sim 16$ days. The FRBs are produced by a highly magnetized pulsar, whose magnetic field is ``combed'' by the strong wind from a companion star, either a massive star or a millisecond pulsar. The FRB pulsar wind retains a clear funnel in the companion's wind that is otherwise opaque to induced Compton or Raman scatterings for repeating FRB emission. The 4 day active window corresponds to the time when the funnel points toward Earth. The interaction also perturbs the magnetosphere of the FRB pulsar and may trigger emission of FRBs. We derive the physical constraints on the comb and the FRB pulsar from the observations and estimate the event rate of FRBs. In this scenario, a lower limit on the period of observable FRBs is predicted. We speculate that both the intrinsic factors (strong magnetic field and young age) and the extrinsic factor (interaction) may be needed to generate FRBs in neutron star binary systems.

%  Comb model for periodic FRB

%  version KI 2020.Jan.30; 2020.Feb.2; Feb. 2b; Feb. 3; Feb. 8; Feb. 17; Feb. 18
\end{abstract}



\section{Introduction}

%Hereafter, we use $Q_x\equiv Q/10^x$ in cgs units.

Fast radio bursts (FRBs) are cosmological radio transients whose origin is enigmatic \citep{Lorimer+07,Thornton+13,Cordes19,Petroff19}.
Regardless of their origin, these bursts can be useful probes for studying cosmology \citep{Ioka03,Inoue04}.

The recent discovery of the periodic repeating FRB 180916.J0158+65 \citep{FRB180916} may bring clues for understanding the source and emission mechanism of repeating FRBs. This source is harbored in a star-forming region of a nearby massive spiral galaxy
at $z=0.0337 \pm 0.0002$, with a
luminosity distance of $149.0 \pm 0.9$ Mpc and a projected size of $\sim 1.5$ kpc \citep{Marcote+20}. 
%star formation rate $>0.016$ M$_{\odot}$ yr$^{-1}$
Twenty-eight bursts were detected from 2018 September 16 to 2019 October 30 by CHIME, which show a period of 
\begin{eqnarray}
  P =16.35\pm 0.18\,{\rm day},
  \label{eq:P}
\end{eqnarray}
with a $\sim 4$ day active time window. The average observed burst rate is\footnote{The true rate would be at least an order of magnitude greater than $\sim 25$ yr$^{-1}$ because CHIME observes the source location a few hours per day and because there could be fainter bursts below the CHIME flux sensitivity.} ${\dot N} \sim 25$ yr$^{-1}$.

%28 bursts from 16 Sep. 2018 to 30 Oct. 2019, ${\dot N} \sim 25$ yr$^{-1}$.

In the literature, magnetars are usually invoked to interpret repeating FRB sources
\citep[e.g.,][]{Popov13,Katz16,Murase16,Kashiyama+17,Kumar+17,Metzger17}.
Alternatively, interaction between an astrophysical stream and neutron star magnetic field
%a neutron star magnetosphere
(the so-called ``cosmic comb'')\footnote{
  In addition to ``interaction,'' we also use ``comb'' because
  it gives the visual impression of the interaction (see Figure~\ref{fig:model}).
}
has been invoked to interpret repeating FRBs \citep{Zhang17,Zhang18}.

Here we propose a binary comb model for the periodic FRB 180916.J0158+65. We interpret the observed period in Equation~(\ref{eq:P}) as the orbital period of a binary system that includes a neutron star for repeating FRBs (the FRB pulsar)\footnote{
  We use ``FRB pulsar'' because the FRB-emitting sources in our model include both young, high-magnetic-field pulsars and magnetars.
}
and a companion whose strong wind imposes a comb on the FRB pulsar.
The interaction
causes both modulation of FRB emission beams and probably also the triggers of FRB emission. We consider the cases that the companion star is either a massive star or a millisecond pulsar. A similar scenario was discussed by \cite{Lyutikov+20}. Alternatively, the $\sim 16$ day period was interpreted as the period of a magnetar due to either free precession \citep{Levin+20,Zanazzi+20} or orbital precession \citep{Yang+20}.
Another model attributes the periodicity to the precession of the jet launched from the accretion disk of a massive black hole \citep{Katz20}.

\section{Physical properties of the binary comb}
\subsection{Binary Separation}

With the observed period (Equation~(\ref{eq:P})) identified as the binary orbital period, the semi-major axis of the binary is
\begin{eqnarray}
  a&=&(GM)^{1/3} \left({P_{\rm orb}}/{2\pi}\right)^{2/3}
  \sim 4 \times 10^{12}\,{\rm cm}\,
  M_{1}^{1/3} P_{{\rm orb},16}^{2/3},
%  \left(\frac{M}{10\,M_{\odot}}\right)^{1/3} \left(\frac{P_{\rm orb}}{16\,{\rm day}}\right)^{2/3},
  \label{eq:a}
\end{eqnarray}
where $M=m_A+m_B = 10 M_{\odot} M_{1}$ is the total mass of the binary
and $P_{{\rm orb},16}=P_{\rm orb}/16$\,days.
The separation between the stars ranges from 
$a_{\min}=a(1-e)$ to $a_{\max}=a(1+e)$ for an eccentricity $e$.
For a massive star companion, the total mass is $M \sim (10$--$50) M_{\odot}$.
For main-sequence stars, the stellar radius is about
$\sim 3\times 10^{11}$\,cm\,$M_{1}^{0.57}$ for $M_1 \gtrsim 0.1$ \citep{Kippenhahn+90},
which is smaller than the binary separation.
%At $P_{\rm orb} \le 20$ days, the observed early-type binaries have eccentricities $0 \le e < 0.4$ \citep{Moe+17}, while the eccentricity may become high after the supernova explosion that makes the FRB pulsar,
%depending on the kick velocity,
%which is small for electron-capture supernovae \citep[e.g.,][]{Postnov+14,Tauris+17}.
For a neutron star companion, the total mass is $M \sim 2.8 M_{\odot}$
and hence $a\sim 2.6 \times 10^{12}$ cm.
%For example, the Galactic sources J0737-3039, J1753-2240, J1811-1736 have $P_{\rm orb}=0.102, 13.638, 18.779$ days and $e=0.088, 0.304, 0.828$, respectively \citep[e.g.,][]{Tauris+17}.

\subsection{Optical Depth}

\subsubsection{Massive Star Companion}

For a massive star case, the wind density around the FRB source is
\begin{eqnarray}
  n_w(0) &\sim& \frac{{\dot M}}{4\pi a^2 m_p V}
  \sim 9\times 10^{5}\,{\rm cm}^{-3}\,
  {\dot M}_{-9} a_{12.6}^{-2} V_{3.3}^{-1},
%  \nonumber\\
%  &\sim& 9\times 10^{5}\,{\rm cm}^{-3}
%  \left(\frac{{\dot M}}{10^{-9}\,M_{\odot}\,{\rm yr}^{-1}}\right)
%  \left(\frac{a}{4\times 10^{12}\,{\rm cm}}\right)^{-2}
%  \nonumber\\
%  &\times&
%  \left(\frac{V}{2\times 10^{3}\,{\rm km}\,{\rm s}^{-1}}\right)^{-1},
\end{eqnarray}
where $a_{12.6}=a/4\times 10^{12}$\,cm,
$V=2\times 10^{3}\,{\rm km}\,{\rm s}^{-1}\,V_{3.3}$ is the wind velocity,
and $m_p$ is the proton mass.
We adopt a mass-loss rate ${\dot M}=10^{-9} M_{\odot}\,{\rm yr}^{-1} M_{-9}$ of main-sequence B stars as the fiducial value
because they are popular (see also Section~\ref{sec:DM}).
Note that the mass-loss rate of O7 and later stars is
a factor of 10--$10^{2}$ lower than theoretically expected \citep{Puls+08,Smith14}.
The B star becomes a rapidly rotating Be star through a mass-exchange episode before the FRB pulsar is born
\citep[e.g.,][]{Postnov+14}.
The equatorial mass-loss rates of Be stars may be larger by a factor of $\sim 10^2$
\citep{Nieuwenhuijzen+88}.

The optical depth to Thomson scattering is small
$\tau_{\rm T} \sim \sigma_{\rm T} n_w(0) a \sim 2\times 10^{-6}$
around the FRB source
where $\sigma_{\rm T}$ is the Thomson cross section.
The optical depth to free-free absorption is
$\tau_{\rm ff} \sim a \alpha_{\nu}^{\rm ff} \sim 0.06\,{\bar g}_{\rm ff}
T_4^{-3/2} \nu_9^{-2}$
%(n_w(0)/9\times 10^{5}\,{\rm cm}^{-3})^{2}$
for fiducial parameters
where
$\alpha_{\nu}^{\rm ff}=(4q^6/3m_ekc)(2\pi/3km_e)^{1/2}T^{-3/2} n_w(0)^2 \nu^{-2} {\bar g}_{\rm ff}$ is the free-free absorption coefficient
at frequency $\nu=1$\,GHz\,$\nu_9$ and temperature $T=10^{4}$\,K\,$T_4$ \citep{Lyutikov+20}.

More important are the induced scattering processes
\citep{Wilson+78,Thompson+94,Lyubarsky08}
because the brightness temperature of the FRB is extremely high 
(e.g., $T_{\rm FRB}\sim 10^{32}$ K) and the scattering probability of bosons is enhanced by the occupation number of the final state
$kT_{\rm FRB}/h\nu \sim 2\times 10^{33}\,T_{{\rm FRB},32}\nu_9^{-1}$.
The optical depth to the induced Compton scattering
at a radius $r=10^{13}\,{\rm cm}\,r_{13}$
is estimated by
\begin{eqnarray}
  \tau_{C} &\sim& \frac{3\sigma_T}{32\pi^2}
  \frac{n_w(r) L c \Delta t}{r^2 m_e \nu^3}
  \sim 30\, {\dot M}_{-9} V_{3.3}^{-1} r_{13}^{-4} (L\Delta t)_{38} \nu_9^{-3},
%  \nonumber\\
%  &\sim& 30
%  \left(\frac{n_w(0)}{9\times 10^{5}\,{\rm cm}^{-3}}\right)
%  \left(\frac{{\dot M}}{10^{-9} M_{\odot}}\right)
%  \left(\frac{V}{2\times 10^{3}\,{\rm km}\,{\rm s}^{-1}}\right)^{-1}
%  \left(\frac{r}{10^{13}\,{\rm cm}}\right)^{-4}
%  \left(\frac{L \Delta t}{10^{38}\,{\rm erg}}\right),
  \label{eq:tauC}
\end{eqnarray}
where $L \Delta t = 10^{38}\,{\rm erg}\,(L\Delta t)_{38}$
is the FRB isotropic luminosity times duration,
and the wind density decreases as $n_w(r) \sim n_w(0) [a/(a+r)]^{2}$
($\propto r^{-2}$ at $r \gg a$).
Here $r$ is measured from the FRB source, not from the companion.
Using a simple criterion for observability $\tau_C<10$
\citep{Lyubarsky08},
the photospheric radius for the induced Compton scatterings is
\begin{eqnarray}
  r_{\rm ph}^{C} &\sim& 1 \times 10^{13}\,{\rm cm}\,
  (L\Delta t)_{38}^{1/4} {\dot M}_{-9}^{1/4} V_{3.3}^{-1/4}
  \nu_9^{-3/4}.
%  \left(\frac{L\Delta t}{10^{38}\,{\rm erg}}\right)^{1/4}
%  \left(\frac{\dot M}{10^{-9} M_{\odot}\,{\rm yr}^{-1}}\right)^{1/4}
%  \nonumber\\
%  &\times&
%  \left(\frac{V}{2\times 10^{3}\,{\rm km}\,{\rm s}^{-1}}\right)^{-1/4}.
  %  \left(\frac{\nu}{1\,{\rm GHz}}\right)^{-3/4},
  \label{eq:rphC}
\end{eqnarray}

The above expression is easy to understand as follows.
The photon occupation number is given by
\begin{eqnarray}
  {\mathscr N} = \frac{c^2 L_{\nu}}{8\pi^2 \theta_b^2 r^2 h \nu^3}
  \sim \frac{c^2 L}{8\pi^2 \theta_b^2 r^2 h \nu^4},
\end{eqnarray}
where $L_{\nu}$ is the isotropic specific luminosity.
For induced Compton scattering, the scattered photon lies within
the half-opening angle of the photon beam $\theta_b$,
so that the cross section is $\sigma_C \sim \sigma_{\rm T} {\mathscr N} \theta_b^2/4$.
In each scattering, a photon loses 
a fraction $\varepsilon_C \sim h\nu \theta_b^2/2 m_e c^2$ of its energy.
Then the effective optical depth is estimated by
$\tau_C \sim \varepsilon_C \sigma_C n_w r$,
which reproduces Equation~(\ref{eq:tauC}) within a factor of $\pi/3$
if we replace $r \theta_b^2/2$ by $c \Delta t$
because the induced scattering occurs
only if the scattered ray remains within the zone
illuminated by the scattering radiation \citep{Lyubarsky08}.
If $\theta_b < (2 c\Delta t/r)^{1/2} \sim 8\times 10^{-3} (\Delta t/10\,{\rm ms})^{1/2} r_{13}^{-1/2}$,
this replacement is not necessary.
Note that the Planck constant $h$ is canceled in the product of ${\mathscr N}$ and $\varepsilon_C$.

The induced Raman scattering by emitting Langmuir waves could be even more significant.
The optical depth at $r=10^{14}\,{\rm cm}\,r_{14}$ is estimated by
\begin{eqnarray}
  \tau_R \sim \tau_C {\nu}/{\nu_p}
  &\sim& 9\,
  {\dot M}_{-9}^{1/2} V_{3.3}^{-1/2} r_{14}^{-3} (L\Delta t)_{38}
  \nu_9^{-2},
%  \left(\frac{{\dot M}}{10^{-9} M_{\odot}}\right)^{1/2}
%  \left(\frac{V}{2\times 10^{3}\,{\rm km}\,{\rm s}^{-1}}\right)^{-1/2}
%  \nonumber\\
%  &\times& \left(\frac{r}{10^{14}\,{\rm cm}}\right)^{-3}
%  \left(\frac{L \Delta t}{10^{38}\,{\rm erg}}\right),
\end{eqnarray}
where $\nu_p=[q^2 n_w(r)/(\pi m_e)]^{1/2}$ is the plasma frequency,
if the scattering angle is not too small and the decay of plasmons is weak
\citep{Thompson+94,Lyubarsky08}.
The photospheric radius for induced Raman scattering is
\begin{eqnarray}
  r_{\rm ph}^{R} &\sim& 1 \times 10^{14}\,{\rm cm}\,
  (L\Delta t)_{38}^{1/3} {\dot M}_{-9}^{1/6} V_{3.3}^{-1/6}
  \nu_9^{-2/3}.
%  \left(\frac{L\Delta t}{10^{38}\,{\rm erg}}\right)^{1/3}
%  \left(\frac{\dot M}{10^{-9} M_{\odot}}\right)^{1/6}
%  \nonumber\\
%  &\times&
%  \left(\frac{V}{2\times 10^{3}\,{\rm km}\,{\rm s}^{-1}}\right)^{-1/6}.
  \label{eq:rphR}
\end{eqnarray}
Note that the Raman scattering effect just widens the beam to
$\theta_b \sim 6\times 10^{-2} (n_w(r)/10^3\,{\rm cm}^{-3})^{1/2} T_4^{-1/2}$
but temporally smears a pulse to $\sim r\theta_b^2/2c > \Delta t$.

The photosphere $r_{\rm ph} \sim 10^{13}$--$10^{14}$ cm
is larger than the separation $a$ in Equation~(\ref{eq:a}) for fiducial parameters.
It is also remarkable that the photosphere is larger than the separation even for a Sun-like star
with $\dot M \sim 2 \times 10^{-14} M_{\odot}$\,yr$^{-1}$,
$M \sim 2.4 M_{\odot}$, and $V \sim 800$ km s$^{-1}$.
Therefore, the stellar wind basically makes the system optically thick.
%Note that we can show that the FRB pulse cannot penetrate the wind
%at the photosphere (see Appendix~\ref{sec:acc}).

\subsubsection{Neutron Star Companion}

For the neutron star companion case, the wind density around the FRB source at
$a \sim 2.6 \times 10^{12}\,{\rm cm}\,a_{12.4}$ is
\begin{eqnarray}
  n_{w}(0) &\sim& \frac{L_w}{4\pi a^2 m_e c^2 V \Gamma (1+\sigma)}
  \sim \frac{5 \times 10^{3}\,{\rm cm}^{-3}}{\Gamma (1+\sigma)}
  L_{w,34} a_{12.4}^{-2},
%  \nonumber\\
%  &\sim& \frac{5 \times 10^{4}\,{\rm cm}^{-3}}{\Gamma (1+\sigma)}
%  \left(\frac{L_w}{10^{35}\,{\rm erg}\,{\rm s}^{-1}}\right)
%  \left(\frac{a}{2.5\times10^{12}\,{\rm cm}}\right)^{-2},
\end{eqnarray}
where we take $V\sim c$,
$\Gamma=[1-(V/c)^2]^{-1/2}$ is the Lorentz factor of the wind,
and $\sigma$ is the ratio of Poynting flux to particle energy flux.
For the fiducial value of the wind luminosity, we take that of a typical millisecond pulsar $L_{w}=10^{34} L_{w,34}$,
because a millisecond pulsar is usually formed in a neutron star binary system
and is the one with the higher spin-down rate as observed in our Galaxy \citep[e.g.,][]{Tauris+17}.

The optical depth to the induced Compton scattering is easy to estimate in the comoving frame of the wind to take relativistic effects into account,
\begin{eqnarray}
  \tau_{C} &\sim& \frac{3\sigma_T}{32\pi^2}\frac{n_{w}'(r) L' c \Delta t'}{r'^2 m_e \nu'^3}
  \nonumber\\
  &\sim& \frac{8 \delta^2}{\Gamma^2(1+\sigma)}
  L_{w,34} r_{12.5}^{-4} (L\Delta t)_{38} \nu_9^{-3},
%  \nonumber\\
%  &\sim& \frac{11}{\delta^2\Gamma^2(1+\sigma)}
%  \left(\frac{L_w}{10^{35}\,{\rm erg}\,{\rm s}^{-1}}\right)
%  \left(\frac{r}{5\times 10^{12}\,{\rm cm}}\right)^{-4}
%  \left(\frac{L \Delta t}{10^{38}\,{\rm erg}}\right),
  \label{eq:tauC2}
\end{eqnarray}
where the relations with the lab-frame quantities are
$L'=L/\delta^4$, $\nu'=\nu/\delta$, $\Delta t'=\delta \Delta t$, $r'=r/\delta$, and $n_{w}'=n_{w}/\Gamma$,
and $\delta=[\Gamma(1-(V/c)\cos\theta_{w})]^{-1}$ is the Doppler factor for an angle $\theta_{w}$
between the photon and the wind direction.
Note that $L/r^2$ transforms as a flux.
%At $\theta_w \lesssim \pi/4$
%as implied by the duty cycle (see Sec.~\ref{sec:duty}),
%the Doppler factor is $\delta \gtrsim \Gamma^{-1}/0.29 $,
%deleting the $\Gamma$ dependence in Eq.~(\ref{eq:tauC2}).
%In this case, the photosphere is located at
The photosphere is located at
\begin{eqnarray}
  r_{\rm ph}^{C} 
  %\lesssim
  \sim
  2\times 10^{12}\,{\rm cm}\,\delta^{1/2} \Gamma^{-1/2} (1+\sigma)^{-1/4} L_{w,34}^{1/4} (L\Delta t)_{38}^{1/4} \nu_9^{-3/4}.
%  \frac{3 \times 10^{12}\,{\rm cm}}{(1+\sigma)^{1/4}}
%  \left(\frac{L_w}{10^{35}\,{\rm erg}\,{\rm s}^{-1}}\right)^{1/4}
%  \left(\frac{L\Delta t}{10^{38}\,{\rm erg}}\right)^{1/4}.
\end{eqnarray}
The photospheric radius for the induced Raman scattering is estimated from $\tau_R \sim \tau_C \nu'/\nu'_p \sim 10$ as
\begin{eqnarray}
  r_{\rm ph}^{R} 
  %\lesssim
  \sim
  3 \times 10^{13}\,{\rm cm}\,\delta^{1/3} \Gamma^{-1/3} (1+\sigma)^{-1/6} L_{w,34}^{1/6} (L\Delta t)_{38}^{1/3} \nu_9^{-2/3}.
%  \frac{4 \times 10^{13}\,{\rm cm}}{\Gamma^{-1/6}(1+\sigma)^{1/6}}
%  \left(\frac{L_w}{10^{35}\,{\rm erg}\,{\rm s}^{-1}}\right)^{1/6}
%  \left(\frac{L\Delta t}{10^{38}\,{\rm erg}}\right)^{1/3}.
\end{eqnarray}
Note that $\delta \sim \Gamma$ for $\theta_w \ll \Gamma^{-1}$, and $\delta \sim 2/(\Gamma \theta_w^2)$ for $\theta_w \gg \Gamma^{-1}$.
%Note that at $\theta_w \lesssim \pi/4$
%as implied by the duty cycle (see Sec.~\ref{sec:duty}),
%the Doppler factor is $\delta \gtrsim \Gamma^{-1}/0.29 $.
Although the Lorentz factor $\Gamma$ and magnetization parameter $\sigma$ are quite uncertain, the dependence is weak, so that the pulsar wind makes the system optically thick
and temporally smears a pulse to $\sim r_{\rm ph}\theta_w^2/2c > \Delta t$ for a large parameter space.
%DO WE NEED THIS SENTENCE? IF SO, PLEASE ADD A REFERENCE FOR THE QUOTED \GAMMA AND \SIGMA VALUES.
%Note that $\Gamma \sim 10^{5}$ and $\sigma \sim 10^{-3}$ reproduce the Crab nebula spectrum,
%while they would be different at different sources and different radii.

For the wind from the FRB pulsar,
the Doppler factor is $\delta \sim \Gamma$ since $\theta_w \sim 0$.
An FRB pulse is likely generated below the photosphere and is temporally smeared to $\sim r_{\rm ph}/2c\Gamma^2$ via scatterings. However, this is shorter than the pulse width $\Delta t$ for large Lorentz factors, and the FRB pulsar itself is observable.
%Then the optical depth in Eq.~(\ref{eq:tauC2}) is small for large Lorentz factors,
%and the FRB pulsar itself is basically transparent.

\subsection{Comb Size Required by Duty Cycle}\label{sec:duty}

\begin{figure}
  \begin{center}
    \includegraphics[width=\linewidth]{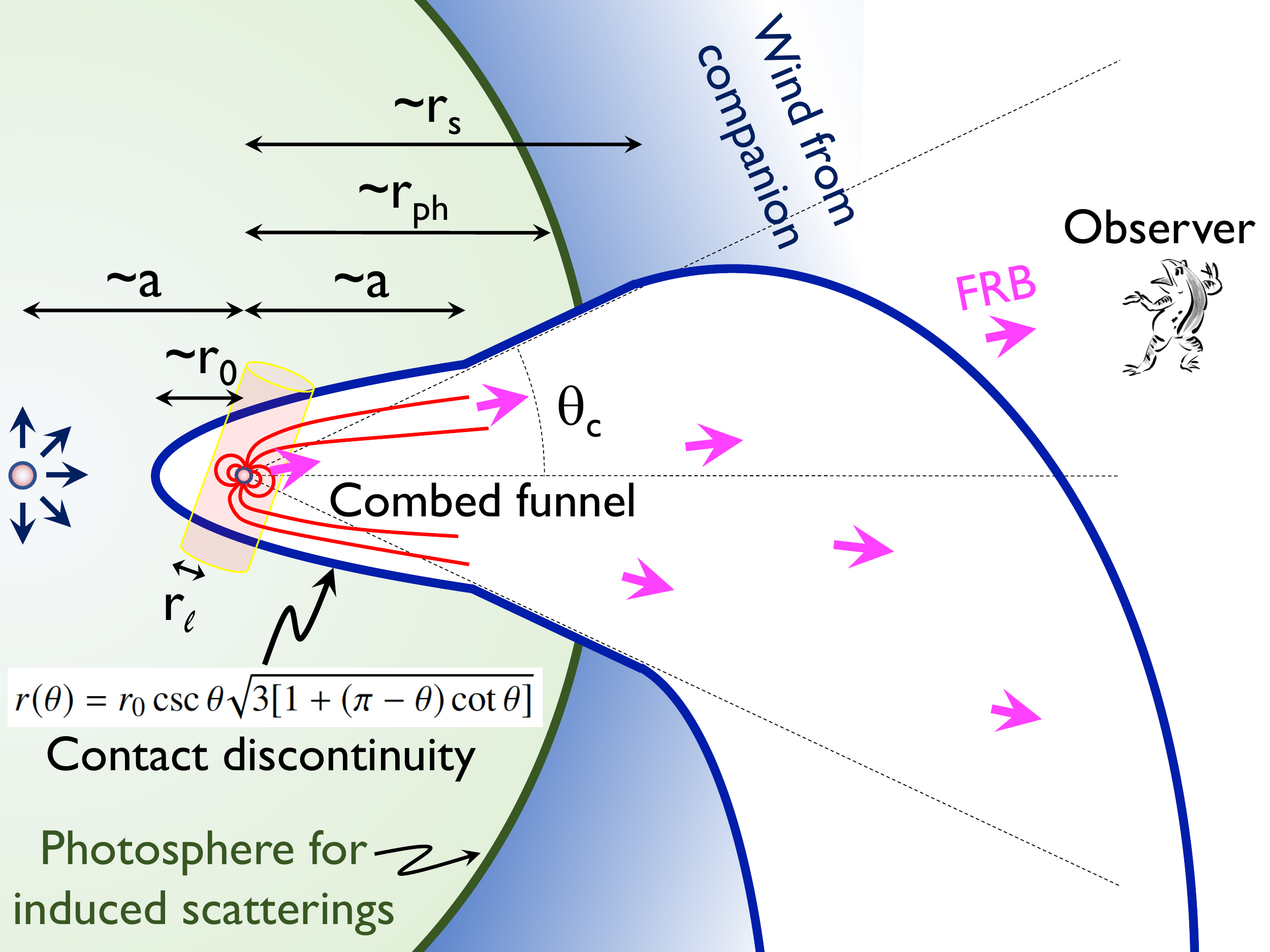}
  \end{center}
  \caption{
    Binary comb model for the periodic FRB 180916.J0158+65. The wind from the FRB pulsar creates a clean funnel with a half-opening angle $\theta_c$, which is combed by the wind from the companion. There are several characteristic scales: $r_{\ell}$ is the radius of the light cylinder, $r_0$ is the minimum comb size on the side of the companion, $a$ is the separation to the companion and is also the radius of the comb opening, $r_{\rm ph}$ is the photospheric radius to induced scatterings, and $r_s$ is the spiral radius due to orbital motion.  
  }
  \label{fig:model}
\end{figure}

The wind from the companion basically makes the system optically thick.
In order to make an FRB observable by an Earth observer,
the wind from the FRB pulsar should open a way to the observer.
Namely, a cosmic comb retains a clear funnel for the FRB to propagate (Figure~\ref{fig:model}).
Since all the bursts arrive in a 4 day phase window of the 16 day period,
the half-opening angle of the comb should be\footnote{If the inclination is close to face-on,
the opening angle should be larger than Equation~(\ref{eq:thc}).
}
\begin{eqnarray}
  \theta_c \gtrsim \pi/4.
  \label{eq:thc}
\end{eqnarray}
%which gives a necessary condition for the wind from the FRB pulsar in the following.
In principle, the opening angle can be arbitrarily small for a highly eccentric orbit
because the polar angle swept by the FRB pulsar during the 4 day phase becomes smaller
for higher eccentricity around the apocenter.
In this case the observable viewing angle is also small.

Near the FRB pulsar with a distance much smaller than the binary separation,
the comb structure is obtained by a problem that 
a wind-blowing star moves with a constant velocity in a uniform density.
The shape of the contact discontinuity is obtained analytically as
\begin{eqnarray}
  r(\theta) = r_0 \csc \theta \sqrt{3[1+(\pi-\theta) \cot\theta]},
\end{eqnarray}
in a thin shock limit \citep{Wilkon96},
where $r(\theta)$ is the distance from the FRB pulsar,
$\theta$ is the polar angle from the axis of symmetry,
and $r_0$ is the minimum size at $\theta=\pi$ (toward the companion).
This solution is consistent with numerical simulations of pulsar bow shocks
\citep{Bucciantini02,Vigelius07}.

The above solution is applicable only up to the binary separation
$r(\theta) \sim a$
because the radial dependence of the companion's wind 
becomes similar to that of the wind from the FRB pulsar
(i.e., the approximation of a uniform density breaks down).
Because of the same radial dependence (e.g., fluxes $\propto r^{-2}$),
the polar angle of the contact discontinuity asymptotically becomes constant,
which determines the opening angle of the comb (Figure~\ref{fig:model}).
Requiring $r(\pi/4) \gtrsim a$ based on Equation~(\ref{eq:thc}), we find that the comb size on the side of the companion should be larger than
\begin{eqnarray}
  r_0 \gtrsim 0.22 a \sim 9 \times 10^{11}\,{\rm cm}\,
  M_1^{1/3} P_{{\rm orb},16}^{2/3}.
%  \left(\frac{M}{10\,M_{\odot}}\right)^{1/3} \left(\frac{P_{\rm orb}}{16\,{\rm day}}\right)^{2/3}.
  \label{eq:r0}
\end{eqnarray}

We stress that the above is a necessary condition.
The duty cycle is also related to the solid angle $\Delta \Omega$
in which the bulk of FRBs are concentrated.
(Note that this is different from
the beaming angle of each FRB $\delta \Omega \sim \pi \theta_b^2$.)
This is even implied by the observations
because the European Very-long-baseline-interferometry Network (EVN)
at 1.7 GHz detected bursts at the leading edge of the activity cycle observed at 400--800 MHz,
while the Effelsberg radio telescope at 1.4 GHz detected no bursts during the middle of the cycle \citep{FRB180916}.
Only a plasma eclipse cannot explain the high-frequency deficit in the middle phase because high-frequency photons are generally transmittable.
However, we should await more observations to confirm the periodicity at high frequencies.

On the other hand, in order for the FRB pulsar to be combed, the companion wind pressure must win at half of the separation, so that the comb size should satisfy
\begin{eqnarray}
  r_0 \lesssim 0.5 a \sim 2 \times 10^{12}\,{\rm cm}\,
  M_1^{1/3} P_{{\rm orb},16}^{2/3}.
%  \left(\frac{M}{10\,M_{\odot}}\right)^{1/3} \left(\frac{P_{\rm orb}}{16\,{\rm day}}\right)^{2/3}.
  \label{eq:r0max}
\end{eqnarray}
Therefore, if the comb triggers FRBs,
the comb size is constrained to a relatively narrow range $0.22 a \lesssim r_0 \lesssim 0.5 a$
in Equations~(\ref{eq:r0}) and (\ref{eq:r0max}).
Remember that the eccentricity relaxes the lower limit while the inclination tightens it.

One more necessary condition arises
because the comb tail is spiraled by the orbital motion at a radius,
\begin{eqnarray}
  r_{s} \sim \frac{V P_{\rm orb}}{2\pi} \frac{\theta_c}{\pi/4}
  \sim 4 \times 10^{13}\,{\rm cm}\,
  P_{{\rm orb},16} V_{3.3} (4\theta_c/\pi),
%  \left(\frac{P_{\rm orb}}{16\,{\rm d}}\right)
%  \left(\frac{V}{2\times 10^{3}\,{\rm km}\,{\rm s}^{-1}}\right)
%  \left(\frac{\theta_c}{\pi/4}\right),
  \label{eq:rs}
\end{eqnarray}
for the massive star case, and
$r_s \sim 7\times 10^{15}$ cm $P_{{\rm orb},16} (4\theta_c/\pi)$
for the neutron star case with $V \sim c$.
This spiral radius should be larger than the photosphere;
otherwise, the wind eventually shields the line of sight
as shown in Figure~\ref{fig:model}.
This condition is marginally satisfied for the massive star case as the photospheric radius is $r_{\rm ph} \sim (10^{13}$--$10^{14})$ cm
in Equations~(\ref{eq:rphC}) and (\ref{eq:rphR}),
while it is satisfied for the neutron star case.
We can also predict a lack of sources with $P_{\rm orb} \lesssim 10$ days for the massive star case and $P_{\rm orb} \lesssim 0.1$ days (with some dependence on $\Gamma$ and $\sigma$) for the neutron star case.

\section{Physical properties of the FRB pulsar}

\subsection{Opacity (Duty Cycle) Constraints}

\begin{figure}
  \begin{center}
    \includegraphics[width=\linewidth]{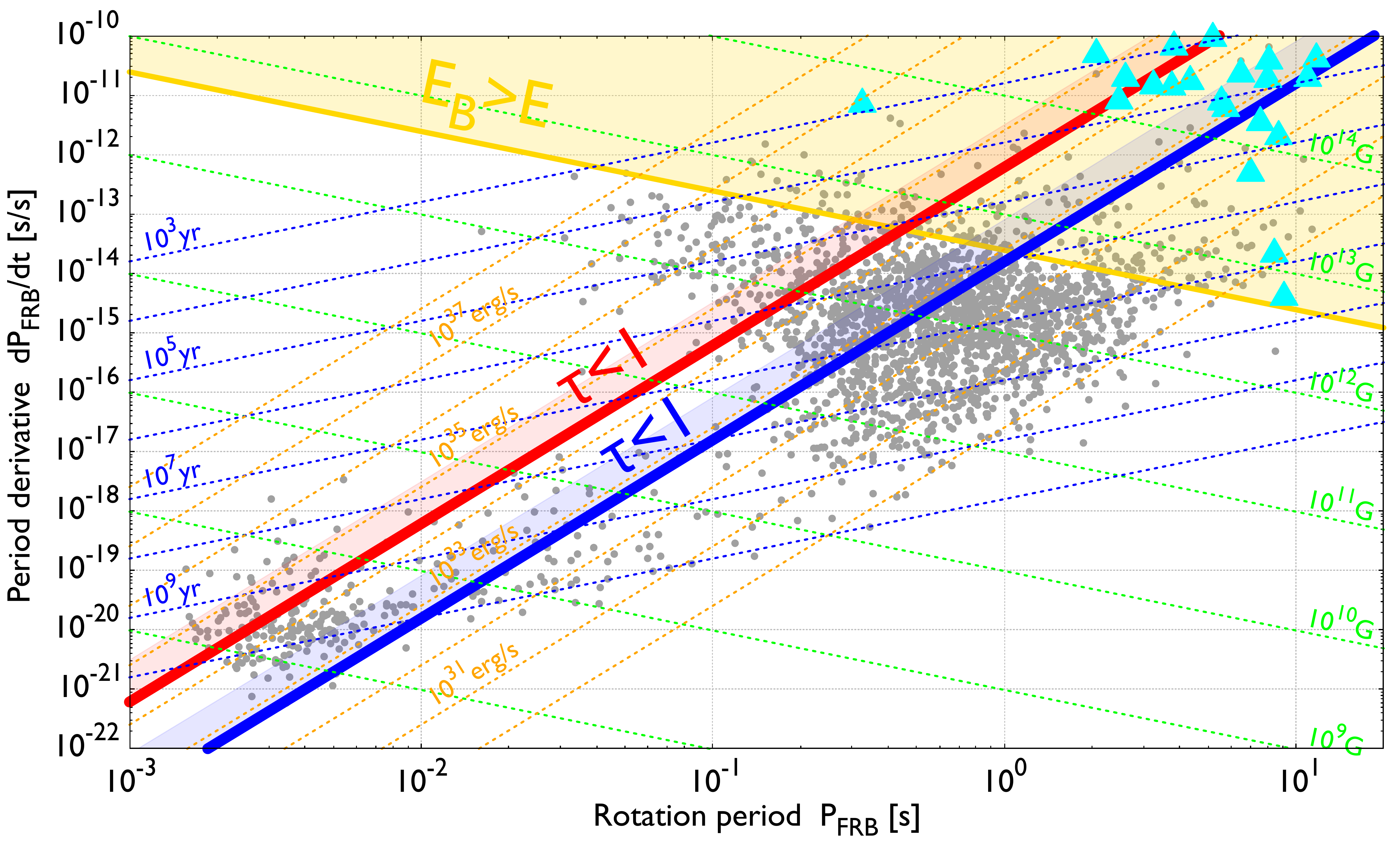}
  \end{center}
  \caption{
    FRB pulsar in $P$--${\dot P}$ diagram, satisfying the opacity constraint in Equation~(\ref{eq:Bp1}) for a massive star companion above the red bold line (where ${\dot P}_{\rm FRB} \propto {\dot M}_{-9}$ for different ${\dot M}$) and in Equation~(\ref{eq:Bp2}) for a neutron star companion above the blue bold line (where ${\dot P}_{\rm FRB} \propto L_{w,34}$ for different $L_w$). In the shaded region, the FRB pulsar is also combed by the wind from the companion. The energy constraint in Equation~(\ref{eq:Bp3}) is satisfied above the yellow solid line. The overlapped region contains magnetars (cyan triangles) and some pulsars (gray dots). Dotted lines show constant lifetime, magnetic field, and spin-down luminosity of pulsars. 
  }
  \label{fig:P-Pdot}
\end{figure}

The comb size necessary for the duty cycle in Equation~(\ref{eq:r0}) is usually larger than
that of the light cylinder of the FRB pulsar with a spin period $P_{\rm FRB}=1\,{\rm s}\,P_{{\rm FRB},0}$,
\begin{eqnarray}
  r_{\ell} \sim c P_{\rm FRB}/(2\pi) \sim 5 \times 10^{9}\, {\rm cm}\,
  P_{{\rm FRB},0}.
%  \left(P_{\rm FRB}/1\,{\rm s}\right).
  \label{eq:rl}
\end{eqnarray}
Then the ram pressure balance between the winds from the companion and the FRB pulsar is expressed by
\begin{eqnarray}
  \frac{L_w}{4\pi a^2 V} \sim \frac{B_p^2}{8\pi}\left(\frac{R}{r_{\ell}}\right)^{6}\left(\frac{r_{\ell}}{r_0}\right)^{2},
\end{eqnarray}
where $R \sim 10$ km is the neutron star radius,
$L_w \sim {\dot M} V^2$ for the massive star case,
and $V\approx c$ for the neutron star case.
For the massive star case, the opacity condition in Equation~(\ref{eq:r0}) gives
\begin{eqnarray}
  B_p \gtrsim 3\times 10^{13}\,{\rm G}\,
  P_{{\rm FRB},0}^2 {\dot M}_{-9}^{1/2} V_{3.3}^{1/2},
%  \left(\frac{P_{\rm FRB}}{1\,{\rm s}}\right)^{2}
%  \left(\frac{\dot M}{10^{-9} M_{\odot}\,{\rm yr}^{-1}}\right)^{1/2}
%  \left(\frac{V}{2\times 10^{3}\,{\rm km}\,{\rm s}^{-1}}\right)^{1/2},
  \label{eq:Bp1}
\end{eqnarray}
where the equality holds if the comb
also stimulates an FRB in Equation~(\ref{eq:r0max}).
For the neutron star case, Equation~(\ref{eq:r0}) gives
\begin{eqnarray}
  B_p \gtrsim 4 \times 10^{12}\,{\rm G}\,
  P_{{\rm FRB},0}^{2} L_{w,34}^{1/2},
%  \left(\frac{P_{\rm FRB}}{1\,{\rm s}}\right)^{2}
%  \left(\frac{L_w}{10^{35}\,{\rm erg}\,{\rm s}^{-1}}\right)^{1/2},
  \label{eq:Bp2}
\end{eqnarray}
where the equality holds if Equation~(\ref{eq:r0max}) is also true.
These conditions are presented in Figure~\ref{fig:P-Pdot}.

\subsection{Energetics Constraints}

The total energy of FRBs during the whole lifetime $t_{\rm life}=10^{4}\,{\rm yr}\,t_{{\rm life},4}$ is
\begin{eqnarray}
  E & =& {\dot N} t_{\rm life} L \Delta t (\Delta \Omega/4\pi) \nonumber\\
  & \sim &
  4 \times 10^{42}\,{\rm erg}\,
  {\dot N}_{25} t_{{\rm life},4} (L\Delta t)_{38} \Delta \Omega_{0.6\pi},
%  \left(\frac{\dot N}{25\,{\rm yr}^{-1}}\right)
%  \left(\frac{t_{\rm life}}{10^{4}\,{\rm yr}}\right)
%  \left(\frac{L\Delta t}{10^{38}\,{\rm erg}}\right)
%  \left(\frac{\Delta \Omega}{0.6\pi}\right),
  \label{eq:E}
\end{eqnarray}
where the observed burst rate is ${\dot N} \sim 25$ yr$^{-1} {\dot N}_{25}$,
the true energy of each burst is smaller by a factor of $\delta \Omega/4\pi$,
and the total number of bursts is increased by a factor of $\Delta\Omega/\delta\Omega$ \citep{Zhang20}.
We take $\Delta \Omega \sim 2\pi (1-\cos(\pi/4)) \sim 0.6\pi$ as a fiducial value
as implied by the duty cycle in Equation~(\ref{eq:thc}).
This energy can be supplied by the magnetic energy $E_{B}>E$ if
\begin{eqnarray}
  B_p \gtrsim 5 \times 10^{12}\,{\rm G}\,
  {\dot N}_{25}^{1/2} t_{{\rm life},4}^{1/2} (L\Delta t)_{38}^{1/2} 
  \Delta \Omega_{0.6\pi}^{1/2}.
%  \left(\frac{\dot N}{25\,{\rm yr}^{-1}}\right)^{1/2}
%  \left(\frac{t_{\rm life}}{10^{4}\,{\rm yr}}\right)^{1/2}
%  \left(\frac{L\Delta t}{10^{38}\,{\rm erg}}\right)^{1/2}
%  \left(\frac{\Delta \Omega}{0.6\pi}\right)^{1/2}.
  \label{eq:Bp3}
\end{eqnarray}
This is marked as the yellow shaded region in Figure~\ref{fig:P-Pdot}.\footnote{Notice that the spin-down luminosity of a pulsar is usually much smaller than the isotropic FRB luminosity \citep[e.g.,][]{Munoz+19}, so that the magnetic energy is most likely the prime mover of FRBs unless the FRB is very narrowly collimated \citep{Katz18}.}

Combining Equations~(\ref{eq:Bp1}), (\ref{eq:Bp2}), and (\ref{eq:Bp3}),
we find the FRB pulsar parameters in the range that includes magnetars and young high-$B$ pulsars
with $B_p \sim 10^{13}$--$10^{15}$ G and $P_{\rm FRB} \sim 1$--$10$ s
%which is thought to have 
for a lifetime $t_{\rm life} \sim 10^{4}$ yr (Figure~\ref{fig:P-Pdot}).
With these parameters,
the FRB pulsar has enough energy
and enough luminosity for pertaining a funnel in the wind from the companion while it is still combed by the wind to trigger FRBs.

Notice that Galactic binary neutron star systems
do not satisfy the energy constraint in Equation~(\ref{eq:Bp3}),
because their ages are much older, i.e., $t_{\rm life} \gg 10^{4}$ yr
\citep[e.g.,][]{Tauris+17}.
Gamma-ray binary systems such as PSR B1259-63 \citep{Aharonian+05}
and PSR J2032+4127 \citep{Lyne+15}
also do not satisfy the energy constraint.
Only a relatively new born pulsar could have
crust or magnetic configurations
that can frequently trigger FRBs via crust cracking or magnetic reconnection.
These intrinsic triggering factors could also explain
why weaker Galactic analogs are not observed.
%just before a reconnection that leads to a FRB.
The pulsar B of the double pulsar system J0737-3039
has $P_{\rm FRB}\sim 2.7735$ s and $B_p=4.9\times 10^{11}$ G.
It also does not satisfy the opacity constraint in Equation~(\ref{eq:Bp2}). 
%Whereas these constraints do not exclude many orders-of-magnitude faint FRBs.
The Galactic high-mass X-ray binaries do not satisfy the opacity constraint because the neutron stars are accreting matter.
Although many Galactic neutron stars satisfy both the opacity and energy constraints, they are not in binaries.
In any case, we do not exclude the possibility that Galactic binary neutron star systems may emit FRB-like signals with much lower luminosities at a much lower rate. Detections or nondetections of such signals in long-term monitoring of these systems may lend support or pose constraints on the scenario proposed in this Letter.

\subsection{FRB Emission}

From the opacity constraint,
the interaction radius in Equation~(\ref{eq:r0})
is larger than the light cylinder in Equation~(\ref{eq:rl}).
The magnetic field at the interaction radius is weaker
than that of the inner magnetosphere near the neutron star surface where
the most energy $E_{B}$ is stored to produce an FRB.
%in Eqs.~(\ref{eq:E}) and (\ref{eq:Bp3}).
Then binary interactions cannot change
the inner magnetic structure to trigger an FRB unless a much tighter binary system is invoked.
%Actually 
In Earth's magnetosphere, for example, magnetic reconnection takes place in the dayside magnetopause and the near-Earth plasma sheet.
No reconnection happens near Earth's surface.

On the other hand, binary interactions may change the current structure (electron density)
in the inner magnetosphere,
which may be related to the coherent condition for FRB emission.
%It is actually observed in pulsars
%such as 
In intermittent pulsars \citep{Kramer+06}
and mode-switching pulsars \citep{Lyne+10},
the change of the spin-down rate
(possibly related to the change of the magnetospheric structure)
%the change of the outer magnetosphere)
is connected with the change of the radio emission condition.
%(i.e., the change of the current near the polar cap).

Based on the energy budget argument, the FRB energy should be dissipated from the inner magnetosphere of the FRB pulsar itself, likely due to crust cracking or magnetic reconnection.
This intrinsic trigger would be similar to that of ordinary magnetars or young, high-magnetic-field pulsars.
However, we speculate that binary interaction may provide the condition to facilitate the FRB coherent radiation mechanism, so that only a small fraction of the intrinsic trigger events can lead to FRBs (see Section~\ref{sec:rate} for more detailed arguments). 
%release is triggered by the FRB pulsar itself
%likely via reconnection,
%while the increase of coherence (i.e., the FRB emission) is triggered by
%the binary interaction.
%Note that 
The interface of the interaction region and the light cylinder of the FRB pulsar makes a connection between the external wind and inner magnetosphere (see Figure~\ref{fig:model}),
%the 
%is a cylinder that can connect
%the interaction region to the inner %magnetosphere
%for electrons to propagate along magnetic field (see Fig.~\ref{fig:model}).
%This is also like 
powering an aurora \citep{Perreault+78,Ebihara+20}
or lightening \citep{Scott+14}
%geomagnetic storm, transient radiation belt
similar to that in Earth's magnetosphere.
Possibly the massive star companion provides
substantial seed photons for creating high-energy particles.
%although the total energy of these phenomena is
%limited by the magnetic energy at the interface.
%In our model, the energy release is triggered by the FRB pulsar itself
%likely via reconnection,
%while the increase of coherence (i.e., the FRB emission) is triggered by
%the binary interaction.
The sudden release of energy in the inner magnetosphere launches a strong particle outflow. When this outflow interacts with the aurora plasma, 
two-stream instability may drive particle bunches that meet the coherent condition for FRBs \citep{Kashiyama+13,Kumar+17,Yang+18}. Alternatively, coherent radiation may be generated by the conversion from reconnection-driven fast magnetosonic waves to electromagnetic waves
\citep{Philippov+19,Lyubarsky20}.
The aurora particles might change the physical parameters such as the conversion radius and Lorentz factor, and hence the coherent condition.
These mechanisms can reproduce the FRB properties such as
polarization and
downward-drifting subpulses
%at a few to tens of MHz ms$^{-1}$
\citep{Wang+19}.

%The coherence trigger may be caused
%by a sudden increase of the wind like a solar flare from the companion,
%while the energy trigger by reconnection is also intermittent.
%These may explain the sporadic nature of the observed FRBs
%\citep{FRB180916}.

%Our model do not specify the FRB radiation mechanism,
%but is compatible with
%bunching coherent curvature radiation
%\citep{Kashiyama+13,Kumar+17,Yang+18}
%or conversion from reconnection-driven fast magnetosonic waves
%to electromagnetic waves
%\citep{Philippov+19,Lyubarsky20}.
%These mechanisms can reproduce the FRB properties such as
%polarization and
%downward-drifting sub-pulses
%\citep{Wang+19}.

\section{Other Constraints}

\subsection{DM, RM, and Persistent Emission}\label{sec:DM}

For FRB 180916.J0158+6, the change of dispersion measure (DM) is constrained as
$\Delta {\rm DM} < 0.1$ pc cm$^{-3}$ \citep{FRB180916}.
This is easily satisfied for the neutron star companion case.
For the massive star companion case,
the contribution to $\Delta {\rm DM}$ only comes from the radius beyond the photosphere.
Considering the difference in the path length during the 4 day active phase, we find that the DM variation is limited by
\begin{eqnarray}
  \Delta {\rm DM} & \lesssim & n_w(r_{\rm ph}) r_{\rm ph} [1-\cos(\pi/4)]
  \nonumber\\
    &\sim& 0.05 \ {\rm pc \ cm^{-3}} \,
%  \sim 0.05\,
  M_{-9} V_{3.3}^{-1} r_{{\rm ph},14}^{-1},
%  \left(\frac{\dot M}{10^{-9} M_{\odot}\,{\rm yr}^{-1}}\right)
%  \left(\frac{V}{2\times 10^{3}\,{\rm km}\,{\rm s}^{-1}}\right)^{-1}
%  \left(\frac{r_{\rm ph}}{10^{14}\,{\rm cm}}\right)^{-1},
\end{eqnarray}
which is consistent with the observation.
More precise observations could detect the DM variation.
Note that at the photosphere,
the electric field of the FRB radiation
is too weak to accelerate electrons to relativistic energies
on the timescale of $(2\pi\nu)^{-1}$ to reduce the DM
\citep{Lu+19,Yuan-Pei+20a}.
The outflow associated with FRBs could also bring additional $\Delta {\rm DM}$
\citep{Yamasaki+19}.

We expect an even larger mass-loss rate for a more massive star
and, hence, a larger $\Delta {\rm DM}$. Thus, a main-sequence B star companion is preferred
for the periodic FRB 180916.J0158+65.
%On the other hand,
About $20$ CHIME bursts do not show a large $\Delta {\rm DM}$,
possibly except for source 5 in \citet{CHIME20}.
Since more massive stars are less abundant, this observation is consistent with our scenario, even though more samples are needed to verify the massive star companion case.
%requiring
%more samples to verify the massive star case.

The rotation measure (RM) of the source is measured as
RM$\sim -114.6 \pm 0.6$ rad m$^{-2}$
\citep{CHIME19b}.
For the neutron star companion case, the expected RM is small.
For the massive star companion case,
there is no evidence of magnetic field for Be stars,
and the dipole field component of 50\% of the stars is
probably weaker than 50 G \citep{Wade+14}.
At the photosphere, this corresponds to
$\lesssim 50\,{\rm G}\, (3\times 10^{11}\,{\rm cm}/r_{\rm ph})^{2}
\sim 5\times 10^{-4}$ G.
The expected absolute value of RM is
$\sim q^3(2\pi m^2 c^4)^{-1} \int_{r_{\rm ph}} n_w B_{\parallel} dr
\lesssim 20$ rad m$^{-2}$, 
less than the observations.
The nondetection of RM variation implies
that the RM comes from a further distance such as the
persistent emission region \citep{Yuan-Pei+20b}.

The persistent radio counterpart is constrained to have a luminosity
$\nu L_{\nu} < 1.3 \times 10^{36}$ erg s$^{-1}$
at 1.7 GHz by the continuum EVN data
and $\nu L_{\nu} < 7.6 \times 10^{35}$ erg s$^{-1}$
at 1.6 GHz by the VLA data \citep{Marcote+20}.
In our model the wind luminosities are less than these constraints.
In the massive star case, the companion mainly shines in the optical band,
also consistent with the observations.

\subsection{Event Rate Density}\label{sec:rate}

Given the burst fluxes $\sim 1$ Jy and
the distance to the periodic FRB $\sim 149$ Mpc
\citep{FRB180916},
a typical luminosity of each burst is $L \sim 10^{40}$ erg s$^{-1}$.
At this luminosity, the event rate density from all the observed FRBs is about
\begin{eqnarray}
  {\cal R}_{\rm FRB}(>L) \lesssim \int_L\phi(L)dL
  \sim 10^{6}\,{\rm Gpc}^{-3}\,{\rm yr}^{-1}
  L_{40}^{-0.8}
%  (L/10^{40}\,{\rm erg}\,{\rm s}^{-1})^{-0.8},
\label{eq:Rfrb}
\end{eqnarray}
by extrapolating and integrating the luminosity function derived from FRBs above $10^{42} \ {\rm erg \ s^{-1}}$ \citep{Luo+20}:
$\phi(L) dL=\phi^{*} \left({L}/{L^{*}}\right)^{\alpha}
  \exp\left(-{L}/{L^*}\right) {dL}/{L^{*}}$
where $\alpha=-1.8$,
$\phi^{*}\sim 339$ Gpc$^{-3}$ yr$^{-1}$
and $L^{*}\sim 2.9 \times 10^{44}$ erg s$^{-1}$. 
Here the ``$\lesssim$'' sign in Equation~(\ref{eq:Rfrb}) denotes the possibility of a luminosity function cutoff below $10^{42} \ {\rm erg \ s^{-1}}$. We note that if this is the case, the argument below is even tighter, so that the current estimate is rather conservative.

If the majority of observed FRBs are dominantly produced by similar sources discussed in this Letter, the true volumetric birth rate of such sources is
\begin{eqnarray}
  {\mathscr R} \sim \frac{{\cal R}_{\rm FRB}}{{\dot N} t_{\rm life}}
  \frac{4\pi}{\Delta \Omega}
  \lesssim
  30\,{\rm Gpc}^{-3}\,{\rm yr}^{-1}\,
  {\dot N}_{25}^{-1} t_{{\rm life},4}^{-1} \Delta \Omega_{0.6\pi}^{-1}.
%  \left(\frac{\dot N}{25\,{\rm yr}^{-1}}\right)^{-1}
%  \left(\frac{t_{\rm life}}{10^{4}\,{\rm yr}}\right)^{-1}
%  \left(\frac{\Delta \Omega}{0.6\pi}\right)^{-1}.
  \label{eq:R}
\end{eqnarray}
This is much smaller than the supernova rate density $\sim 10^5$ Gpc$^{-3}$ yr$^{-1}$
and the magnetar birth rate density $\sim 10^4$ Gpc$^{-3}$ yr$^{-1}$ \citep{Kaspi+17}
for a reasonable lifetime $t_{\rm life} \gg 1$ yr.
The expected number of the FRB sources is only $\lesssim 0.03 {\dot N}_{25}^{-1} \Delta \Omega_{0.6\pi}^{-1}$ in our Galaxy.
This suggests that a special condition is necessary to limit the amount of sources to produce FRBs
and binary interaction is an attractive solution.
Thus, our model is not like a single magnetar model. Binary interaction is an important ingredient of the model.

For the massive star companion case,
the reduction factors from the supernova rate density include
the fraction of pulsars that satisfies the FRB condition (may be comparable to the magnetar fraction) by $\sim 0.1$ \citep{Kaspi+17},
the fraction of the right binary separation
by $\sim 0.1$ \citep{Moe+17},
the survival fraction of the kick
at the first supernova by $\sim 0.1$ \citep{Postnov+14,Tauris+17},
the mass ratio and so on by $\sim 0.3$ \citep{Moe+17}.
Thus, it is natural to have a small birth rate similar to Equation~(\ref{eq:R}).

For the neutron star companion case,
the birth rate of binary neutron stars
with a separation of $a \sim 10^{12}$ cm is estimated as
$\sim 10^{2}$ Gpc$^{-3}$ yr$^{-1}$ by the population synthesis \citep{Belczynski+02}, which is
smaller by a factor of $\sim 10$ than the merger rate derived from gravitational-wave observations
\citep{GW170817,GW190425}.
This is also consistent with the number of
Galactic binary neutron star systems \citep{Tauris+17}.
By multiplying the magnetar fraction $\sim 0.1$,
it also gives a small birth rate similar to Equation~(\ref{eq:R}).

\section{Summary and discussions}\label{sec:summary}

We have shown that if the periodicity of FRB 180916.J0158+65  is due to the binary period, the interaction between a young, strongly magnetized neutron star (the FRB pulsar) and a companion with a strong wind can give the right conditions to interpret the observations. 
The wind from the companion makes the system optically thick to FRB photons
due to induced Compton or Raman scatterings.
The FRB pulsar should make a clear funnel by blowing a wind, which is preserved by a cosmic comb of the FRB pulsar magnetic field. 
The production of FRBs requires that the FRB pulsar be young and highly magnetized. 
%field close to that of magnetars. 
%which requires the parameter range of a magnetar.https://ja.overleaf.com/project/5e4b9f1d9bbfd900016b7d19
Since supernova explosions
%magnetars 
provide too high a birth rate density that may overproduce repeating FRBs,
a special condition may be required to facilitate FRB production.
We suggest that
interactions in binary systems may be essential to generate FRBs besides the intrinsic conditions (strong magnetic fields and young age) imposed on the FRB pulsars.
%triggers FRBs by increasing the coherence in the emission.

We predict a lack of FRBs with $P_{\rm orb}\lesssim 10$ days
for a massive star companion and
$P_{\rm orb}\lesssim 0.1$ days (with some dependence on $\Gamma$ and $\sigma$) for a neutron star companion,
because the funnel is spiraled by the orbital motion within the photosphere in those cases.
We also suggest that a DM variation is close to the detection limit
for the massive star case, requiring more samples with the precision of measurements.

The periodic FRB 180916.J0158+65 is apparently different from
the first repeater FRB 121102,
which emits bright FRBs and is accompanied
by a bright persistent source $\sim 10^{39}$ erg s$^{-1}$
and high $|{\rm RM}| \sim 10^{5}$ rad m$^{-2}$
\citep{Spitler+16,Chatterjee+17,Marcote+17,Tendulkar+17}.
That source may imply a different type of companion
such as a supermassive black hole \citep{Zhang18}.\footnote{
  Possible periodic activity was also reported for FRB 121102
  \citep{Rajwade+20}
  after the submission of this Letter.
}

There might exist genuinely non-repeating FRBs that comprise a small fraction of the total FRB population, but with distinct emission properties
\citep[e.g., non-repeating FRB pulses appear to be narrower than repeating ones;][]{CHIME19b,CHIME20}. These bursts may have catastrophic origins
such as neutron star mergers \citep{Totani13}, white dwarf mergers \citep{Kashiyama+13}, and so on.

\acknowledgments

The authors would like to thank the anonymous referee for useful comments, and
H. Gao, W. Ishizaki, K. Kashiyama, S. Kisaka, P. Kumar, K. Kyutoku, K. Murase, N. Seto, S. Shibata, Y. Suwa, S. Tanaka, and C. Thompson for useful discussions.
This research made use of  the ATNF pulsar catalog
\citep[\url{https://www.atnf.csiro.au/research/pulsar/psrcat/};][]{Manchester+05} and McGill Online Magnetar Catalog
\citep[\url{http://www.physics.mcgill.ca/~pulsar/magnetar/main.html};][]{Olausen+14}.
This work is partly supported by
JSPS KAKENHI Nos. 18H01215, 17H06357, 17H06362, 17H06131, 26287051
(KI).
%by the Grant-in-Aid from the Ministry of Education, Culture, Sports,
%Science and Technology (MEXT) of Japan.
Discussions during the YITP workshop YITP-T-19-04 and YKIS2019
were also useful to complete this work.
B.Z. acknowledges UNLV for granting a sabbatical leave and YITP for hospitality and support.

\bibliography{ref}{}
\bibliographystyle{aasjournal}



\end{document}